\documentclass[12pt]{iopart}
\usepackage{graphicx}  
\begin{document}
\def\NZ{{\bar N_0}}
\def\NJ{N_{\J}}
\def\J{J/\psi}
\def\Nc{N_c}
\def\Ncbar{N_{\bar c}}
\def\ccbar{c \bar c}
\def\Nccbar{N_{\ccbar}}
\def\dNJ{d\NJ/dy}

\title{An assessment of $\J$ formation in the light of initial RHIC data}

\author{R. L. Thews}  

\address{Department of Physics, University of Arizona, Tucson, AZ 85721, USA}

\begin{abstract}
Predictions of $\J$ formation at RHIC via ``off-diagonal" combinations
of charm and anticharm quarks in a region of color deconfinement are confronted
with initial data from the PHENIX collaboration.  We find that the 
measured centrality behavior places significant constraints on the
various parameters which control model calculations of $\J$ formation.
Within present statistical and systematic uncertainties, one can map out
a region of parameter space within which 
the contribution of formation in a deconfined phase 
is allowed.  As these uncertainties decrease and new data from d-Au 
interactions becomes available, it is expected that definitive tests
for the presence of this formation mechanism will be possible.  We
anticipate that the rapidity and transverse momentum spectra will
prove decisive for a final determination.

\end{abstract}




\section{Introduction}
Initial data at RHIC energy on $\J$ production in Au-Au collisions
has been eagerly awaited, in terms of a signal for the presence of
color deconfinement \cite{Matsui:1986dk}.
Of special interest is the possibility that
a direct extrapolation of anomalous suppression from the SPS energy
range could be supplanted by a new formation mechanism fueled by 
the presence of multiple pairs of charm quarks in each nuclear
collision at sufficiently high energy \cite{Thews:2000rj}.
  One can argue on general grounds
that the resulting $\J$ formation will grow quadratically with
the total number of unbound charm quark pairs, with a normalization factor 
which depends on the specific formation mechanism \cite{Thews:2001hy}.  
Predictions for
this type of formation have been made for two specific models.  
The statistical hadronization model \cite{Braun-Munzinger:2000px}
assumes that at the time of hadronization, charm quarks
are distributed into hadrons according to statistical rules, incorporating 
an additional fugacity factor to conserve charm.  The kinetic formation
model \cite{Thews:2002jg, Thews:2001em} 
considers $\J$ formation within the region of deconfinement, 
and calculates the net number remaining at hadronization due to
a competition between formation and breakup reactions.  In this work, we
confront the kinetic formation model predictions with the initial
data from the PHENIX collaboration at RHIC.  This initial data 
\cite{Frawley:2002vz, Nagle:2002ib} consists
of a measurement of $\J$ produced at central rapidity
 in Au-Au collisions at $\sqrt s$ = 200 GeV, presented 
in three centrality bins.  
In addition, there is data on $\J$ production
in pp collisions in three rapidity bins.  
There is also an indirect measurement of initial charm production via
detection of high-$p_t$ electrons \cite{Averbeck:2002nz}.

In this presentation we
first give in the next section a brief summary of the
kinetic model formulation, with a discussion of primary
uncertainties and ranges of parameters.  The following section
is devoted to a comparison of the {\em predictions} of this
model with the initial PHENIX numbers, both the absolute
values and the centrality dependence.  Finally, we 
explore the regions of parameter space which are consistent
with the central values of the initial data, and comment on
aspects for the future.

\section{Kinetic formation model}

The kinetic formation model is most easily motivated in 
a scenario where suppression of $\J$ due to deconfined color is
related to their breakup via collisions with gluons \cite{Kharzeev:1994pz}.  
Given 
the distribution of gluons and the cross section for this
reaction, one can calculate the rate for destruction of 
$\J$ in the region of deconfinement.  It is then clear that
one should also take into account the inverse reaction, in which
charm and anticharm quarks interact to form $\J$ with emission
of a gluon.  The net effect is a competition between these
two processes, which one can express via a Boltzmann
equation for the charm quark and $\J$ populations.

\begin{equation}\label{eqkin}
\frac{d\NJ}{dt}=
  \lambda_{\mathrm{F}} N_c\, \rho_{\bar c } -
    \lambda_{\mathrm{D}} \NJ\, \rho_g\,,
\end{equation}
where 
 $\rho$ denotes
number density,
and the reactivity $\lambda$ is
the reaction rate $\langle \sigma v_{\mathrm{rel}} \rangle$
averaged over the momentum distribution of the initial
participants, i.e. $c$ and $\bar c$ for $\lambda_F$ and
$\J$ and $g$ for $\lambda_D$.
In our calculations, this equation is solved numerically while
enforcing exact charm conservation.  At SPS energy, one finds that
the formation process is negligble, since the average number of charm
quark pairs produced even in central collisions is much smaller
than unity.  At high energy where one expects the number
of initial charm pairs $\Nccbar$ will become large, one can find an approximate analytic
solution for Eq. \ref{eqkin} which exhibits the anticipated
quadratic dependence.

\begin{equation}
\NJ(t) = \epsilon(t) \times  [\NJ(t_0) + 
\Nccbar^2 \int_{t_0}^{t}
{\lambda_{\mathrm{F}}\, [V(t^{\prime})\, \epsilon(t^{\prime})]^{-1}\, dt^{\prime}}],
\label{eqbeta}
\end{equation}
where
$\epsilon(t) = e^{-\int_{t_0}^{t}{\lambda_{\mathrm{D}}\, \rho_g\,
dt}}$
would be the equilvalent suppression factor in this scenario if the
formation mechanism were neglected.
Here we have assumed the densities are uniform within a
deconfinement volume $V(t)$.  We use a thermal gluon density, initial
temperature $T_0$ as a variable parameter, and fixed final deconfinement 
temperature.  The deconfinement volume is assumed to
expand isentropically, for scenarios in which transverse expansion is
controlled by a parameter vtr.  The initial population of $\J$ is taken
to be a fraction $x$ of the initial number of charm quark pairs.  The cross
section is taken from an OPE-based model, based on the color dipole interaction
of a nonrelativistic bound state with a coulomb bound state spectrum.  For the centrality dependence, we utilize a calculation of the number of participants
as a function of impact parameter.  
We also calculate the participant
density in the transverse plane, and define an effective transverse area
as the ratio of total to density.  The participant density is also
used to specify the centrality dependence of initial temperature $T_0$. 
All centrality-dependent quantities are then scaled to their b=0 values.
\begin{figure}[htb]
\begin{center}
\includegraphics[clip=,width=10.5cm,clip]{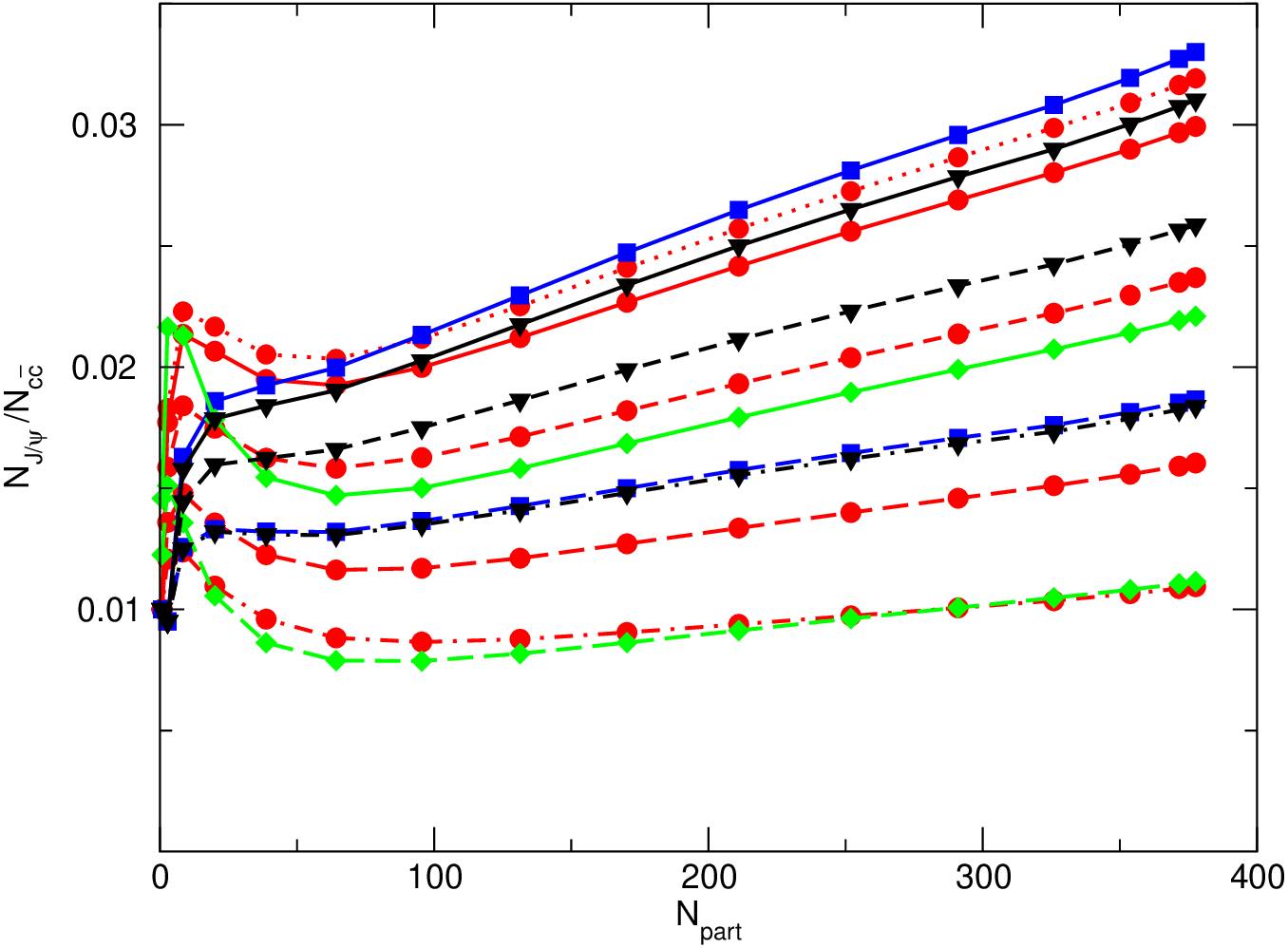}
\caption{Predictions for $\J$ formation. All use $\Nccbar$ = 10 and x = 0.01 .}
\label{kineticpredict1}
\end{center}
\end{figure}
\vskip -0.7cm
Our initial calculations were performed for $T_0$ = 0.3, 0.4, and 0.5 GeV,
x = 0.01, and $\Nccbar$ (b=0) = 10.  To minimize the dependence on 
$\Nccbar$, the ratio $\NJ$ / $\Nccbar$ was presented.  
Shown in Figure \ref{kineticpredict1} are predictions for this ratio as published
in Ref. \cite{Thews:2000rj}, where the legends are defined.  The primary 
sensitivity was due to the charm quark momentum distribution. As might
be expected, a thermal distribution was most efficient in formation, while
distributions with various increasing rapidity widths predicted smaller
numbers.  All of the curves rise with increasing centrality, and
have magnitudes near or above the assumed initial value (x = 0.01).    

\section{PHENIX data vs. predictions}

\begin{figure}[htb]
\begin{center}
\includegraphics[clip=,width=10.5cm]{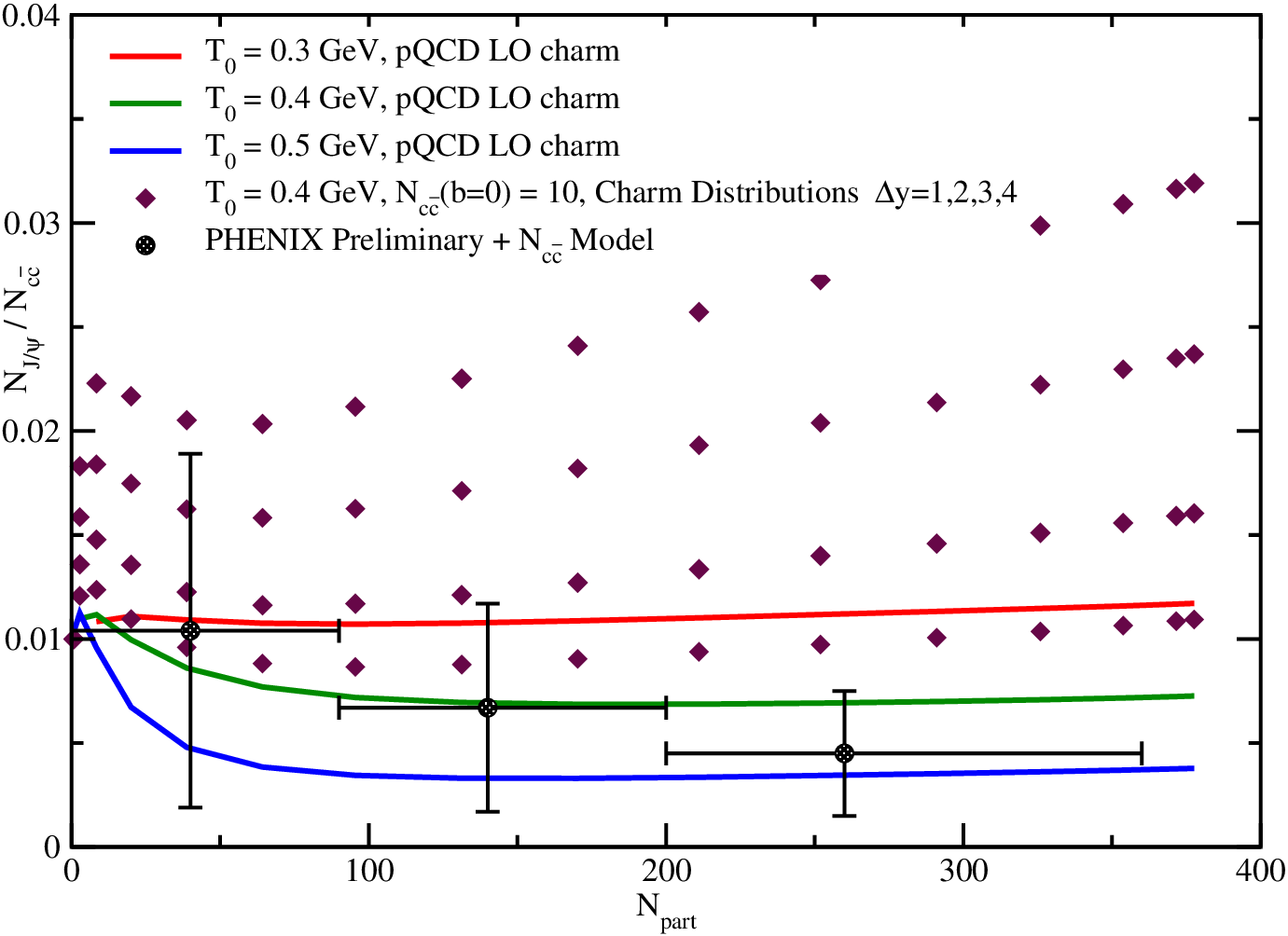}
\caption{Indirect comparison of formation model with PHENIX data.}
\label{kineticpredict2}
\end{center}
\end{figure}

For an initial comparison of these predictions with the PHENIX data, one 
must translate the measured $\dNJ$ at y=0 to the total $\NJ$, and
calculate the initial $\Nccbar$ in each centrality bin.  For the former,
we assumed a flat $\J$ rapidity distribution with an effective $\Delta y$ = 4.
The $\Nccbar$ for each centrality region were scaled to the 
$\Nccbar$(b=0) = 10 assumed in the calculated curves, and varied with
centrality according to the nuclear overlap function $T_{AA}$(b) evaluated
at an impact parameter corresponding to the central values of the
experimental bins.  The calculated ratios obtained in this way from the
data are shown in Figure \ref{kineticpredict2}, where the error bars
are taken from a sum of statistical and systematic uncertainties.  One
sees that the central values decrease with centrality.  (This is the
same behavior as shown in Reference \cite{Frawley:2002vz} for the
measured $\J$ rapidity density scaled by the number of binary collisions.) 
It must be remembered that the ratios extracted from the experimental data in
this way are only valid if the experimental value of total charm production
proves to be the same as that assumed in these model calculations. 
For ease of comparison, one set of model predictions for $T_0$ = 0.4 GeV and 
charm quark momentum widths $\Delta y$ = 1,2,3,4 are shown from among
those in Figure \ref{kineticpredict1}.  Also shown are three additional
model curves which use the exact charm quark momentum distribution from a
LO pQCD calculation. 
\vskip -0.7cm
 These latter results are in general agreement with
the preliminary data.  For a more systematic comparison of the data with
formation model results, we will restrict the charm momentum distributions
to the pQCD set.

\section{PHENIX data and formation parameters}
\begin{figure}[htb]
\begin{center}
\includegraphics[clip=,width=10.5cm]{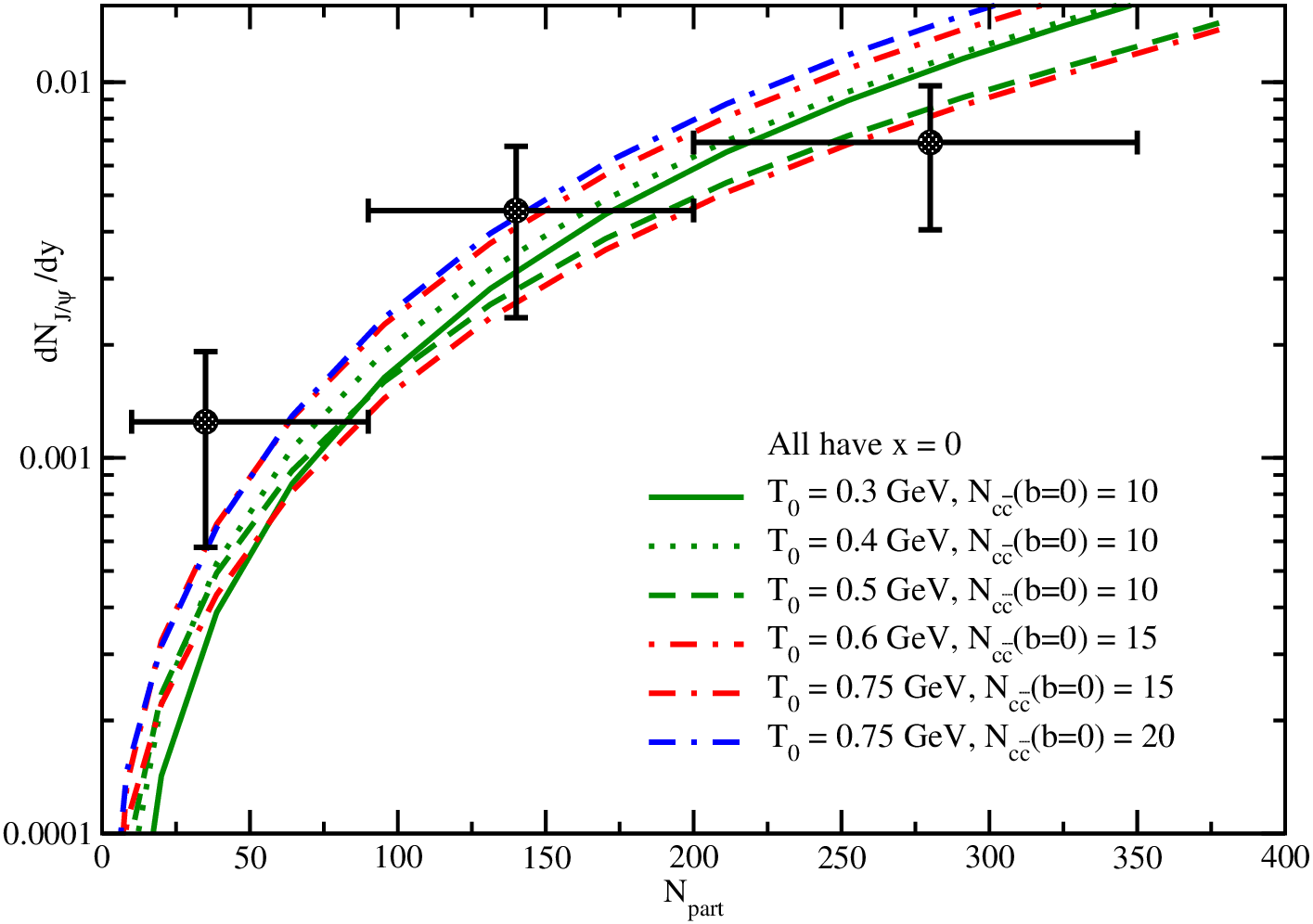}
\caption{Kinetic model variation with $T_0$ and $\Nccbar$.}
\label{parameterscanvt0}
\end{center}
\end{figure}

\begin{figure}[htb]
\begin{center}
\includegraphics[clip=,width=10.5cm]{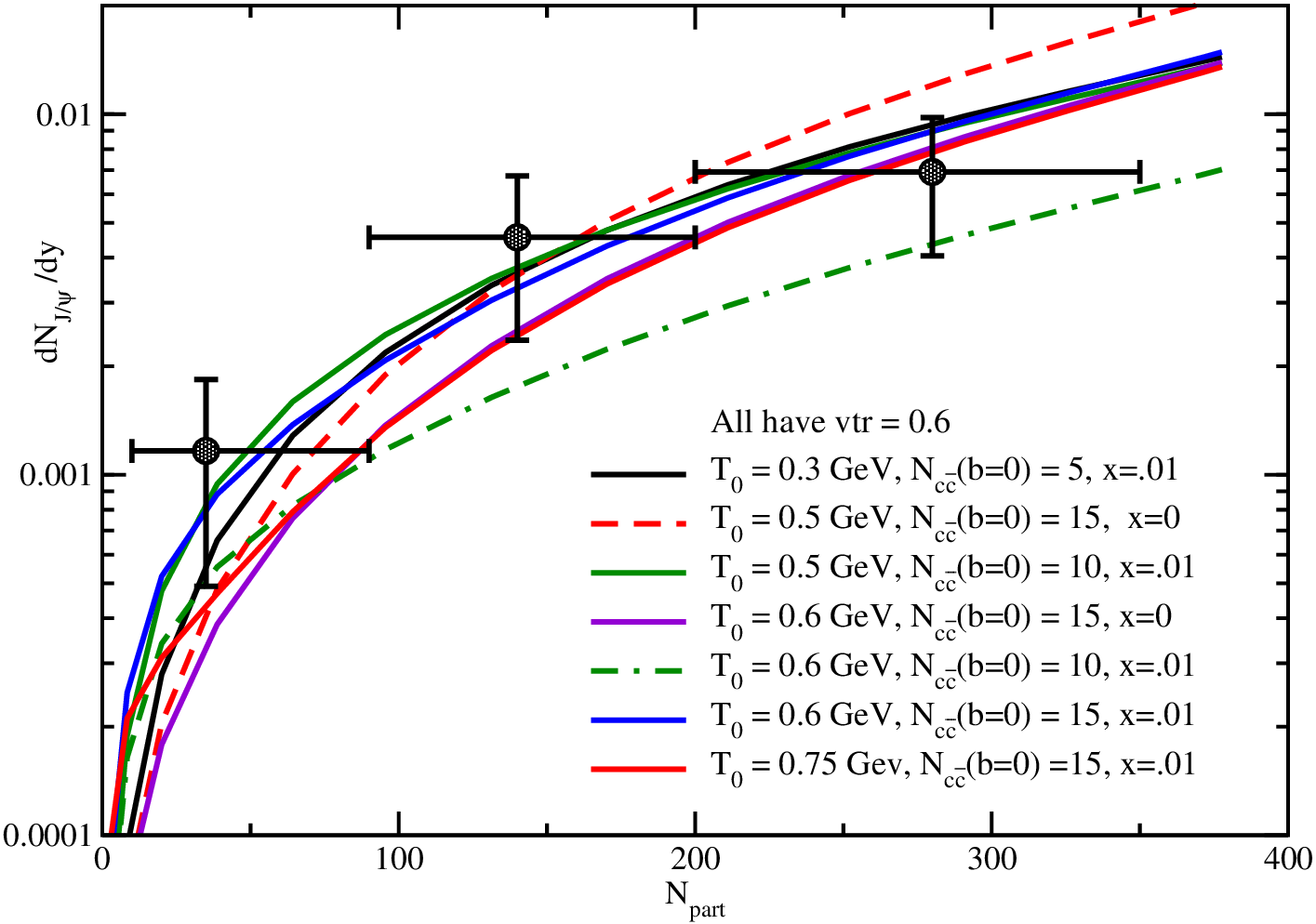}
\caption{Kinetic model results for d$\NJ$/dy with transverse expansion.}
\label{parameterscanvt6}
\end{center}
\end{figure}

\begin{figure}[htb]
\begin{center}
\includegraphics[clip=,width=10.5cm]{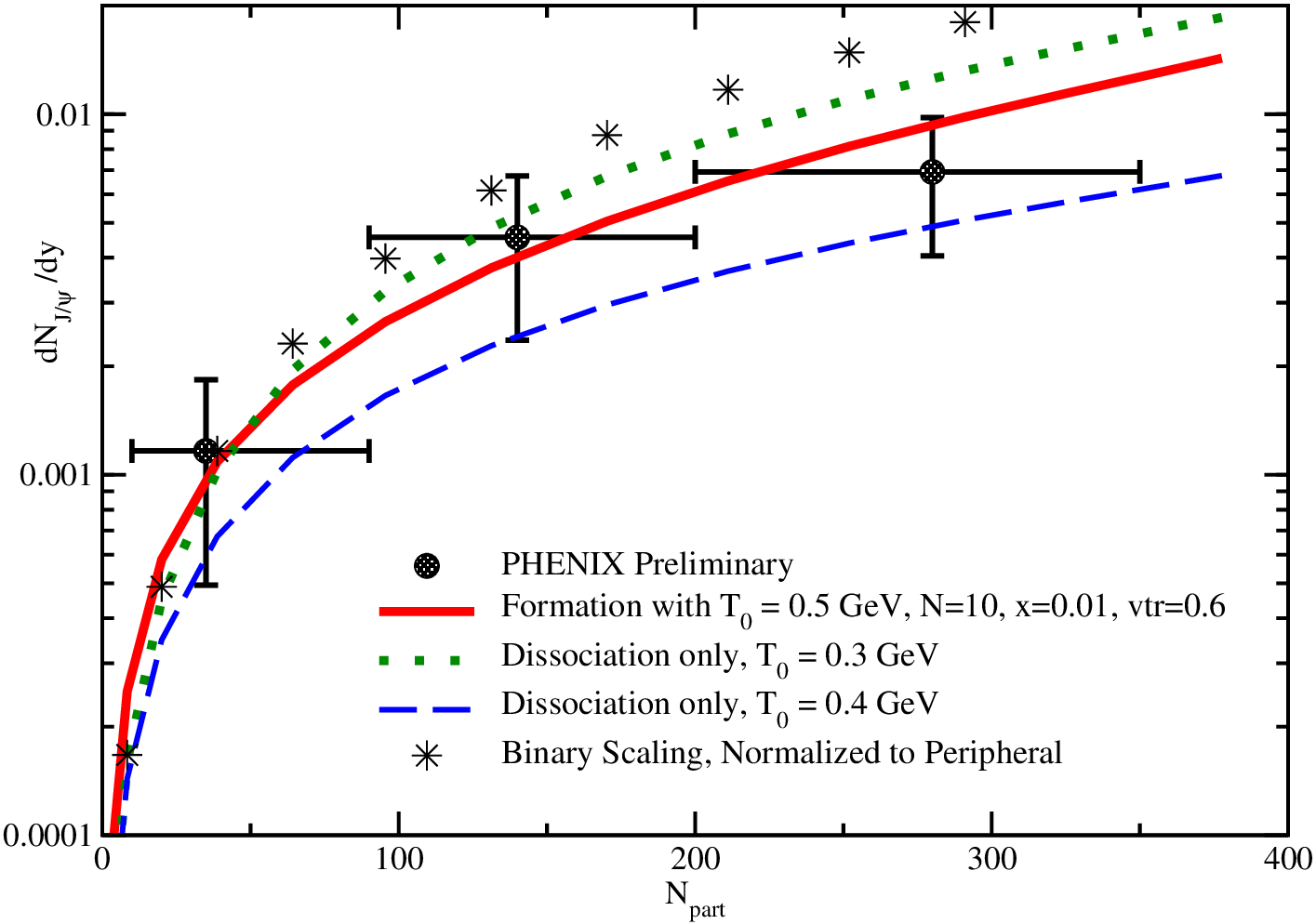}
\caption{Comparison of formation formation calculation and binary scaling with PHENIX data.}
\label{dnjpsiplusbinary}
\end{center}
\end{figure}

In this section we adapt the model calculations to compare directly 
with $\dNJ$ data.  The calculated $\NJ$ have one component originating
from initial production and a second from the combined formation/dissociation
processes.  The first is converted to a rapidity density according to
the measured $\J$ rapidity distribution from production in pp interactions 
\cite{Frawley:2002vz}, and the second from a model calculation of the
rapidity distribution which follows from the formation mechanism 
\cite{thewsinprep}.  One also needs the $\Nccbar$ values, which we 
estimate from preliminary open charm cross section \cite{Averbeck:2002nz}.
Since there is presently an uncertainty of order factor of two in
this estimate, we choose to present calculations for a wide range
of $\Nccbar$(b=0) = 5,10,15,20.  We start with zero for our initial production
parameter $x$.
Given these
parameter constraints, we scan over a range of $T_0$ to select
sets which are roughly compatible with the data.  We show in 
Figure \ref{parameterscanvt0} such a set.
One may question how the formation model is able to produce
these curves, which rise with centrality less rapidly than
binary scaling in order to be compatible with the data.  The
answer is that the formation model
involves an inverse volume, which would normally provide a linear
$N_{part}$ dependence.  However this volume is time dependent and
is integrated over the deconfinement lifetime along with factors
sensitive to the dissociation probability.  We also extend the
parameter space to include a nonzero transverse expansion of the
deconfinement volume, shown in Figure 
\ref{parameterscanvt6}.
One sees that there is a significant correlation between sets of
allowed parameters, and that a precise experimental constraint on
$\Nccbar$, for example, would place significant constraints on the
others.  
\vskip -0.7cm
\vskip -0.7cm
Finally, we show for completeness an alternate scenario in 
Figure \ref{dnjpsiplusbinary}.  
Along with the data and one set of formation model parameters which 
reproduces the central values of the measurements, we show two solutions
in which only the dissociation mechanism is nonzero, using 
$T_0$
= 0.3 GeV and 0.4 GeV which
control the gluon density and the deconfinement lifetime.
For comparison, the centrality dependence which would result
from pure binary collision scaling are also shown.
It appears that even if the uncertainties in the data points were reduced
considerably, one could probably find an acceptable  dissociation 
fit.  
In that case, it will be essential to compare not only the
magnitudes and centrality dependence, but also to confront  the
different scenarios with measured $\J$
rapidity and transverse momentum behavior.  Work along these lines is
underway.\\
\vskip 0.3cm
{\bf Acknowledgment}: This work was supported in part by 
U.S. Department of Energy
Grant DE-FG03-95ER40937.

\end{document}